# Interactive Analytical Processing in Big Data Systems: A Cross-Industry Study of MapReduce Workloads

Yanpei Chen, Sara Alspaugh, Randy Katz
University of California, Berkeley
{ychen2, alspaugh, randy}@eecs.berkeley.edu

## ABSTRACT

Within the past few years, organizations in diverse industries have adopted MapReduce-based systems for large-scale data processing. Along with these new users, important new workloads have emerged which feature many small, short, and increasingly interactive jobs in addition to the large, long-running batch jobs for which MapReduce was originally designed. As interactive, large-scale query processing is a strength of the RDBMS community, it is important that lessons from that field be carried over and applied where possible in this new domain. However, these new workloads have not yet been described in the literature. We fill this gap with an empirical analysis of MapReduce traces from six separate business-critical deployments inside Facebook and at Cloudera customers in e-commerce, telecommunications, media, and retail. Our key contribution is a characterization of new MapReduce workloads which are driven in part by interactive analysis, and which make heavy use of query-like programming frameworks on top of MapReduce. These workloads display diverse behaviors which invalidate prior assumptions about MapReduce such as uniform data access, regular diurnal patterns, and prevalence of large jobs. A secondary contribution is a first step towards creating a TPC-like data processing benchmark for MapReduce.

## 1. INTRODUCTION

Many organizations depend on MapReduce to handle their large-scale data processing needs. As companies across diverse industries adopt MapReduce alongside parallel databases [5], new MapReduce workloads have emerged that feature many small, short, and increasingly interactive jobs. These workloads depart from the original MapReduce use case targeting purely batch computations, and shares semantic similarities with large-scale interactive query processing, an area of expertise of the RDBMS community. Consequently, recent studies on query-like programming extensions for MapReduce [14, 27, 49] and applying query optimization techniques to MapReduce [16, 23, 26, 31, 34, 43] are likely to bring considerable benefit. However, integrating these ideas into business-critical systems requires configuration tuning and performance benchmarking against real-life production MapReduce workloads. Knowledge of such workloads is currently limited to a handful of technology companies [8,11,17,38,41,48]. A cross-workload comparison is thus far absent, and use cases beyond the technology industry have not been described. The increasing diversity of MapReduce operators create a pressing need to characterize industrial MapReduce workloads across multiple companies and industries.

Arguably, each commercial company is rightly advocating for their particular use cases, or the particular problems that their products address. Therefore, it falls to neutral researchers in academia to facilitate cross-company collaboration, and mediate the release of cross-industries data.

In this paper, we present an empirical analysis of seven industrial MapReduce workload traces over long-durations. They come from production clusters at Facebook, an early adopter of the Hadoop implementation of MapReduce, and at e-commerce, telecommunications, media, and retail customers of Cloudera, a leading enterprise Hadoop vendor. Cumulatively, these traces comprise over a year's worth of data, covering over two million jobs that moved approximately 1.6 exabytes spread over 5000 machines (Table 1). Combined, the traces offer an opportunity to survey emerging Hadoop use cases across several industries (Cloudera customers), and track the growth over time of a leading Hadoop deployment (Facebook). We believe this paper is the first study that looks at MapReduce use cases beyond the technology industry, and the first comparison of multiple large-scale industrial MapReduce workloads.

Our methodology extends [17–19], and breaks down each MapReduce workload into three conceptual components: data, temporal, and compute patterns. The key findings of our analysis are as follows:

- There is a new class of MapReduce workloads for interactive, semi-streaming analysis that notably differs from the original use case targeting purely batch computations.
- There is a wide range of behavior within this workload class, such that we must exercise caution in regarding any aspect of workload dynamics as "typical".
- Query-like programatic frameworks on top of MapReduce such as Hive and Pig make up a considerable fraction of activity in all workloads we analyzed.
- Some prior assumptions about MapReduce such as uniform data access, regular diurnal patterns, and prevalence of large jobs no longer hold.

Subsets of these observations have emerged in several studies that each looks at only one MapReduce workload [11, 14, 18, 19, 27, 49]. Identifying these characteristics across a rich and diverse set of workloads shows that the observations are applicable to a range of use cases.

We view this class of MapReduce workloads for interactive, semi-streaming analysis as a natural extension of interactive query processing. Their prominence arises from the ubiquitous ability to generate, collect, and archive data about both technology and physical systems [24], as well as the growing statistical literacy across many industries to interactively explore these datasets and derive timely insights [5, 14, 33, 39]. The semantic proximity of this MapReduce workload to interactive query processing suggests that optimization techniques for one likely translate to the other, at least in principle. However, the diversity of behavior even within this same MapReduce workload class complicates efforts to develop generally applicable improvements. Consequently, ongoing MapReduce studies that draw on database management insights would benefit from checking workload assumptions against empirical measurements.

The broad spectrum of workloads analyzed allows us to identify the challenges associated with constructing a TPC-style big data processing benchmark for MapReduce. Top concerns include the complexity of generating representative data and processing characteristics, the lack of understanding about how to scale down a production workload, the difficulty of modeling workload characteristics that do not fit well-known statistical distributions, and the need to cover a diverse range of workload behavior.

The rest of the paper is organized as follows. We review prior work on workload-related studies (§ 2) and develop hypotheses about MapReduce behavior using existing mental models. We then describe the MapReduce workload traces (§ 3). The next few sections present empirical evidence that describe properties of MapReduce workloads for interactive, semi-streaming analysis, which depart from prior assumptions about MapReduce as a mostly batch processing paradigm. We discuss data access patterns (§ 4), workload arrival patterns (§ 5), and compute patterns (§ 6). We detail the challenges these workloads create for building a TPC-style benchmark for MapReduce (§ 7), and close the paper by summarizing the findings, reflecting on the broader implications of our study, and highlighting future work (§ 8).

## 2. PRIOR WORK

The desire for thorough system measurement predates the rise of MapReduce. Workload characterization studies have been invaluable in helping designers identify problems, analyze causes, and evaluate solutions.

Workload characterization for database systems culminated in the TPC-* series of benchmarks [51], which built on industrial consensus on representative behavior for transactional processing workloads. Industry experience also revealed specific properties of such workloads, such as Zipf distribution of data accesses [28], and bimodal distribution of query sizes [35]. Later in the paper, we see that some of these properties also apply to the MapReduce workloads we analyzed.

The lack of comparable insights for MapReduce has hindered the development of a TPC-like MapReduce benchmark suite that has a similar level of industrial consensus and representativeness. As a stopgap alternative, some MapReduce microbenchmarks aim to faciliate performance comparison for a small number of large-scale, stand-alone jobs [4, 6, 45], an approach adopted by a series of studies [23, 31, 34, 36]. These microbenchmarks of stand-alone jobs remain different from the perspective of TPC-* benchmarks, which views a *workload* as a complex superposition of many jobs of various types and sizes [50].

The workload perspective for MapReduce is slowly emerging, albeit in point studies that focus on technology industry use cases one at a time [8, 11, 12, 38, 41, 48]. The stand-alone nature of these studies forms a part of an interesting historical trend for workload-based studies in general. Studies in the late 1980s and early 1990s capture system behavior for only one setting [37, 44], possibly due to the nascent nature of measurement tools at the time. Studies in the 1990s and early 2000s achieve greater generality [13, 15, 25, 40, 42, 46], likely due to a combination of improved measurement tools, wide adoption of certain systems, and better appreciation of what good system measurement enables. Stand-alone studies have become common again in recent years [8, 11, 17, 38, 41, 48], likely the result of only a few organizations being able to afford large-scale systems.

The above considerations create the pressing need to generalize beyond the initial point studies for MapReduce workloads. As MapReduce use cases diversify and (mis)engineering opportunities proliferate, system designers need to optimize for common behavior, in addition to improving the particulars of individual use cases.

Some studies amplified their breadth by working with ISPs [25, 46] or enterprise storage vendors [13], i.e., intermediaries who interact with a large number of end customers. The emergence of enterprise MapReduce vendors present us with similar opportunities to look beyond single-point MapReduce workloads.

### 2.1 Hypotheses on Workload Behavior

One can develop hypotheses about workload behavior based on prior work. Below are some key questions to ask about any MapReduce workload.

1. For optimizing the underlying storage system:
   - How uniform or skewed are the data accesses?
   - How much temporal locality exists?

2. For workload-level provisioning and load shaping:
   - How regular or unpredictable is the cluster load?
   - How large are the bursts in the workload?

3. For job-level scheduling and execution planning:
   - What are the common job types?
   - What are the size, shape, and duration of these jobs?
   - How frequently does each job type appear?

4. For optimizing query-like programming frameworks:
   - What % of cluster load comes from these frameworks?
   - What are the common uses of each framework?

5. For performance comparison between systems:
   - How much variation exists between workloads?
   - Can we distill features of a representative workload?

Using the original MapReduce use case of data indexing in support of web search [22] and the workload assumptions behind common microbenchmarks of stand-alone, large-scale jobs [4, 6, 45], one would expect answers to the above to be: (1). Some data access skew and temporal locality exists, but there is no information to speculate on how much. (2). The load is sculpted to fill a predictable web search diurnal with batch computations; bursts are not a concern since new load would be admitted conditioned on spare cluster capacity. (3). The workload is dominated by large-scale jobs with fixed computation patterns that are repeatedly and regularly run. (4). We lack information to speculate how and how much query-like programming frameworks are used. (5). We expect small variation between different use cases, and the representative features are already captured in publications on the web indexing use case and existing microbenchmarks.

Several recent studies offered single use case counter-points to the above mental model [11, 14, 18, 19, 27, 49]. The data in this paper allow us to look across use cases from several industries to identify an alternate workload class. What surprised us the most is (1). the tremendous diversity within this workload class, which precludes an easy characterization of representative behavior, and (2). that some aspects of workload behavior are polar opposites of the original large-scale data indexing use case, which warrants efforts to revisit some MapReduce design assumptions.

## 3. WORKLOAD TRACES OVERVIEW

We analyze seven workloads from various Hadoop deployments. All seven come from clusters that support business-critical processes. Five are workloads from Cloudera's enterprise customers in e-commerce, telecommunications, media, and retail. Two others are Facebook workloads on the same cluster across two different time periods. These workloads offer an opportunity to survey Hadoop use cases across several technology and traditional industries (Cloudera customers), and track the growth of a leading Hadoop deployment (Facebook).

Table 1 provides details about these workloads. The trace lengths are limited by the logistical feasibility of shipping the trace data for offsite analysis. The Cloudera customer workloads have raw logs approaching 100GB, requiring us to set up specialized file transfer tools. Transferring raw logs is infeasible for the Facebook workloads, requiring us to query Facebook's internal monitoring tools. Combined, the workloads contain over a year's worth of trace data, covering a significant amount of jobs and bytes processed by the clusters.

The data comes from standard logging tools in Hadoop; no additional tools were necessary. The workload traces contain per-job summaries for job ID (numerical key), job name (string), input/shuffle/output data sizes (bytes), duration, submit time, map/reduce task time (slot-seconds), map/reduce task counts, and input/output file paths (string). We call each of the numerical characteristic a *dimension* of a job. Some traces have some data dimensions unavailable.

We obtained the Cloudera traces by doing a time-range selection of per-job Hadoop history logs based on the file timestamp. The Facebook traces come from a similar query on Facebook's internal log database. The traces reflect no logging interruptions, except for the cluster in `CC-d`, which was taken offline several times due to operational reasons.

| Trace | Machines | Length | Date | Jobs | Bytes moved |
|---|---|---|---|---|---|
| CC-a | <100 | 1 month | 2011 | 5759 | 80 TB |
| CC-b | 300 | 9 days | 2011 | 22974 | 600 TB |
| CC-c | 700 | 1 month | 2011 | 21030 | 18 PB |
| CC-d | 400-500 | 2+ months | 2011 | 13283 | 8 PB |
| CC-e | 100 | 9 days | 2011 | 10790 | 590 TB |
| FB-2009 | 600 | 6 months | 2009 | 1129193 | 9.4 PB |
| FB-2010 | 3000 | 1.5 months | 2010 | 1169184 | 1.5 EB |
| Total | >5000 | ≈ 1 year | - | 2372213 | 1.6 EB |

Table 1: Summary of traces. CC is short for "Cloudera Customer". FB is short for "Facebook". Bytes moved is computed by sum of input, shuffle, and output data sizes for all jobs.

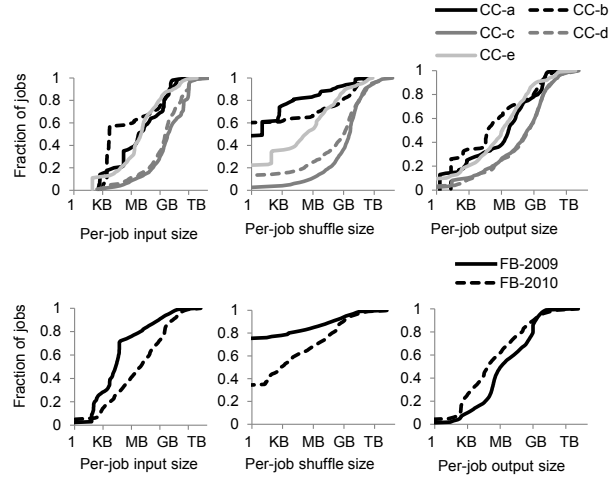

Figure 1: Data size for each workload. Showing input, shuffle, and output size per job.

There are some inaccuracies at trace start and termination, due to partial information for jobs straddling the trace boundaries. The length of our traces far exceeds the typical job length on these systems, leading to negligible errors. To capture weekly behavior for `CC-b` and `CC-e`, we intentionally queried for 9 days of data to allow for inaccuracies at trace boundaries.

## 4. DATA ACCESS PATTERNS

Data manipulation is a key function of any data management system, so understanding data access patterns is crucial. Query size, data skew, and access temporal locality are key concerns that impact performance for RDBMS systems. The mirror considerations exist for MapReduce. Specifically, this section answers the following questions:
– How uniformly or skewed are the data accesses?
– How much temporal locality exists?
We begin by looking at per job data sizes, the equivalent of query size (§ 4.1), skew in access frequencies (§ 4.2), and temporal locality in data accesses (§ 4.3).

### 4.1 Per-job Data Sizes

Figure 1 shows the distribution of per-job input, shuffle, and output data sizes for each workload. Across the workloads, the median per-job input, shuffle, and output sizes differ by 6, 8, and 4 orders of magnitude, respectively. Most jobs have input, shuffle, and output sizes in the MB to GB range. Thus, benchmarks of TB and above [4,6,45] captures only a narrow set of input, shuffle, and output patterns.

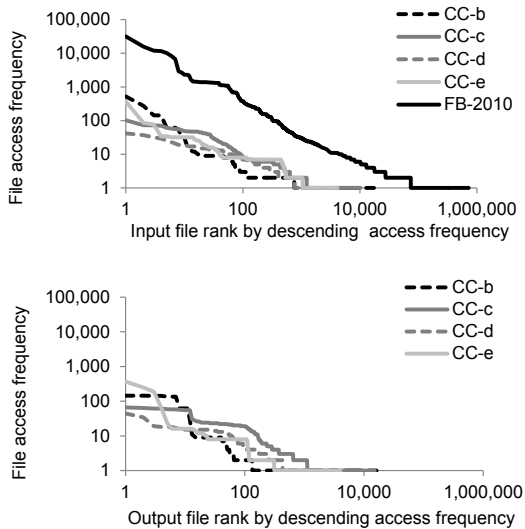

**Figure 2: Log-log file access frequency vs. rank. Showing Zipf distribution of same shape (slope) for all workloads.**

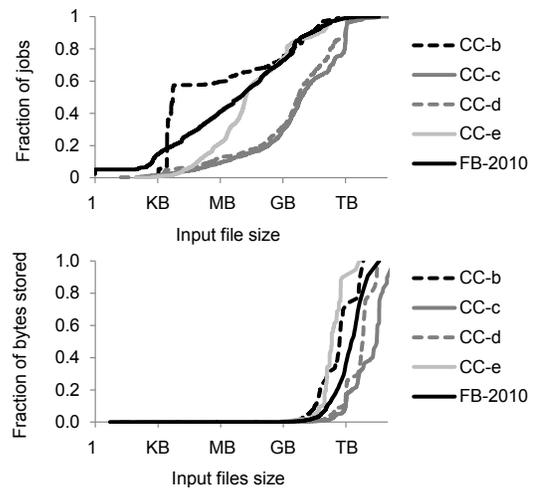

**Figure 3: Access patterns vs. input file size. Showing cummulative fraction of jobs with input files of a certain size (top) and cummulative fraction of all stored bytes from input files of a certain size (bottom).**

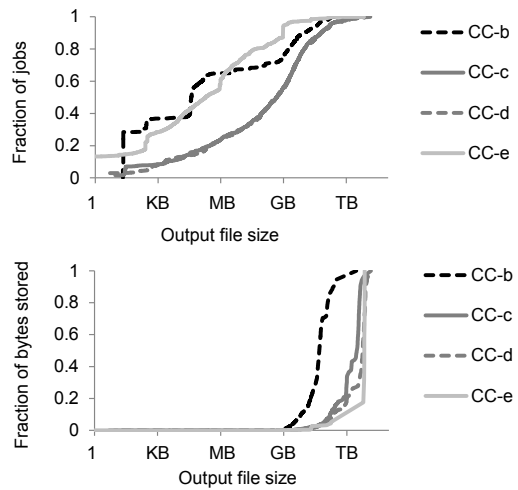

**Figure 4: Access patterns vs. output file size. Showing cummulative fraction of jobs with output files of a certain size (top) and cummulative fraction of all stored bytes from output files of a certain size (bottom).**

From 2009 to 2010, the Facebook workloads' per-job input and shuffle size distributions shift right (become larger) by several orders of magnitude, while the per-job output size distribution shifts left (becomes smaller). Raw and intermediate data sets have grown while the final computation results have become smaller. One possible explanation is that Facebook's customer base (raw data) has grown, while the final metrics (output) to drive business decisions have remained the same.

## 4.2 Skews in Access Frequency

This section analyzes HDFS file access frequency and intervals based on hashed file path names. The `FB-2009` and `CC-a` traces do not contain path names, and the `FB-2010` trace contains path names for input only.

Figure 2 shows the distribution of HDFS file access frequency, sorted by rank according to non-decreasing frequency. Note that the distributions are graphed on log-log axes, and form approximately straight lines. This indicates that the file accesses follow a Zipf-like distribution, i.e., a few files account for a very high number of accesses. This observation challenges the design assumption in HDFS that all data sets should be treated equally, i.e., stored on the same medium, with the same data replication policies. Highly skewed data access frequencies suggest a tiered storage architecture should be explored [12], and *any* data caching policy that includes the frequently accessed files will bring considerable benefit. Further, the slope parameters of the distributions are all approximately 5/6 across workloads and for both inputs and outputs. Thus, file access patterns are Zipf-like distributions of the same shape. Figure 2 suggests the existence of common computation needs that lead to the same file access behavior across different industries.

The above observations indicate only that caching helps.

If there is no correlation between file sizes and access frequencies, maintaining cache hit rates would require caching a fixed fraction of bytes stored. This design is not sustainable, since caches intentionally trade capacity for performance, and cache capacity grows slower than full data capacity. Fortunately, further analysis suggests more viable caching policies.

Figures 3 and 4 show data access patterns plotted against input and output file sizes. The distributions for fraction of jobs versus file size vary widely (top graphs), but converge in the upper right corner. In particular, 90% of jobs accesses files of less than a few GBs (note the log-scale axis). These files account for up to only 16% of bytes stored (bot-tom graphs). Thus, a viable cache policy is to cache files whose size is less than a threshold. This policy allows cache capacity growth rates to be detached from the growth rate in data.

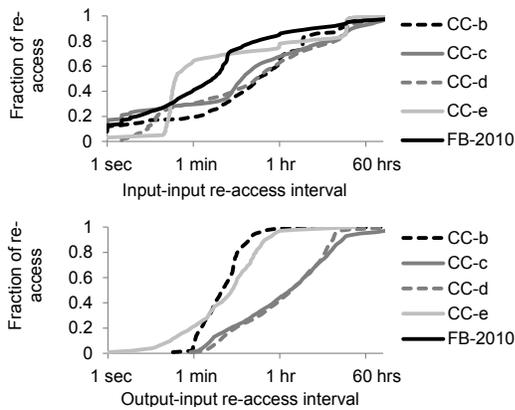

**Figure 5: Data re-accesses intervals. Showing interval between when an input file is re-read (top), and when an output is re-used as the input for another job (bottom).**

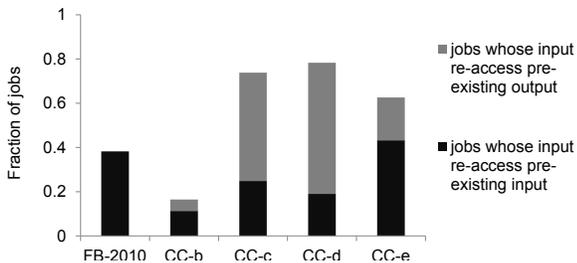

**Figure 6: Fraction of jobs that reads pre-existing input path. Note that output path information is missing from `FB-2010`.**

Prior work has also observed Zipf-like distributed data access patterns for RDBMS workloads, culminating in the formulation of the 80-20 rule, i.e., 80% of the data access go to 20% of the data [28]. For MapReduce, the rule is more complicated. We need to consider both the input and output data sets, and the size of each data set. If we had just considered a Zipf log-log slope of 5/6, we would have arrived at a 80-40 rule. Figure 3 and 4 account for the size of data sets also, and indicate that 80% of jobs (data accesses) go to less than 10% of the stored bytes, for both input and output data sets. Depending on the workload, the access patterns range from an 80-1 rule to an 80-8 rule.

### 4.3 Access Temporal Locality

Further analysis also reveals temporal locality in the data accesses. Figure 5 indicates the distribution of time intervals between data re-accesses. 75% of the re-accesses take place within 6 hours. Thus, a possible cache eviction policy is to evict entire files that have not been accessed for longer than a workload specific threshold duration. Any similar policy to least-recently-used (LRU) would make sense.

Figure 6 further shows that up to 78% of jobs involve data re-accesses (`CC-c`, `CC-d`, `CC-e`), while for other work-loads, the fraction is lower. Thus, the same cache eviction policy potentially translates to different benefits for different workloads.

Combined, the observations in this section indicate that it will be non-trivial to preserve for performance comparisons the data size, skew in access frequency, and access temporal locality of the data. The analysis also reveals the tremendous diversity across workloads. Only one numerical feature remains relatively fixed across workloads—the shape parameter of the Zipf-like distribution for data access frequencies. Consequently, we should be cautious in considering any aspect of workload behavior as being "typical".

## 5. WORKLOAD VARIATION OVER TIME

The temporal workload intensity variation has been an important concern for RDBMS systems, especially ones that back consumer-facing systems subject to unexpected spikes in behavior. The transactions or queries per second metric quantifies the maximum stress that the system can handle. The analogous metric for MapReduce is more complicated, as each job or "query" in MapReduce potentially involves different amounts of data, and different amounts of computation on the data. Actual system occupancy depends on the combination of these multiple time-varying dimensions, with thus yet unknown correlation between the dimensions.

The empirical workload behavior over time has implications for provisioning and capacity planning, as well as the ability to do load shaping or consolidate different workloads. Specifically, this section tries to answer the following:
— How regular or unpredictable is the cluster load?
— How large are the bursts in the workload?

In the following, we look at workload variation over a week (§ 5.1), quantify burstiness, a common feature for all workloads (§ 5.2), and compute temporal correlations between different workload dimensions (§ 5.3). Our analysis proceeds in four dimensions — the job submission counts, the aggregate input, shuffle, and output data size involved, the aggregate map and reduce task times, and the resulting system occupany in the number of active task slots.

### 5.1 Weekly Time Series

Figure 7 depicts the time series of four dimensions of workload behavior over a week. The first three columns respectively represents the cumulative job counts, amount of I/O (again counted from MapReduce API), and computation time of the jobs submitted in that hour. The last column shows cluster utilization, which reflects how the cluster serviced the submitted workload described by the preceding columns, and depends on the cluster hardware and execution environment.

Figure 7 shows all workloads contain a high amount of noise in all dimensions. As neither the signal nor the noise models are known, it is challenging to apply standard signal processing methods to quantify the signal to noise ratio of these time series. Further, even though the number of jobs submitted is known, it is challenging to predict how much I/O and computation will result.

Some workloads exhibit daily diurnal patterns, revealed by Fourier analysis, and for some cases, are visually identifiable (e.g., jobs submission for `FB-2010`, utilization for `CC-e`). In Section 7, we combine this observation with several others to speculate that there is an emerging class of interactive and semi-streaming workloads.

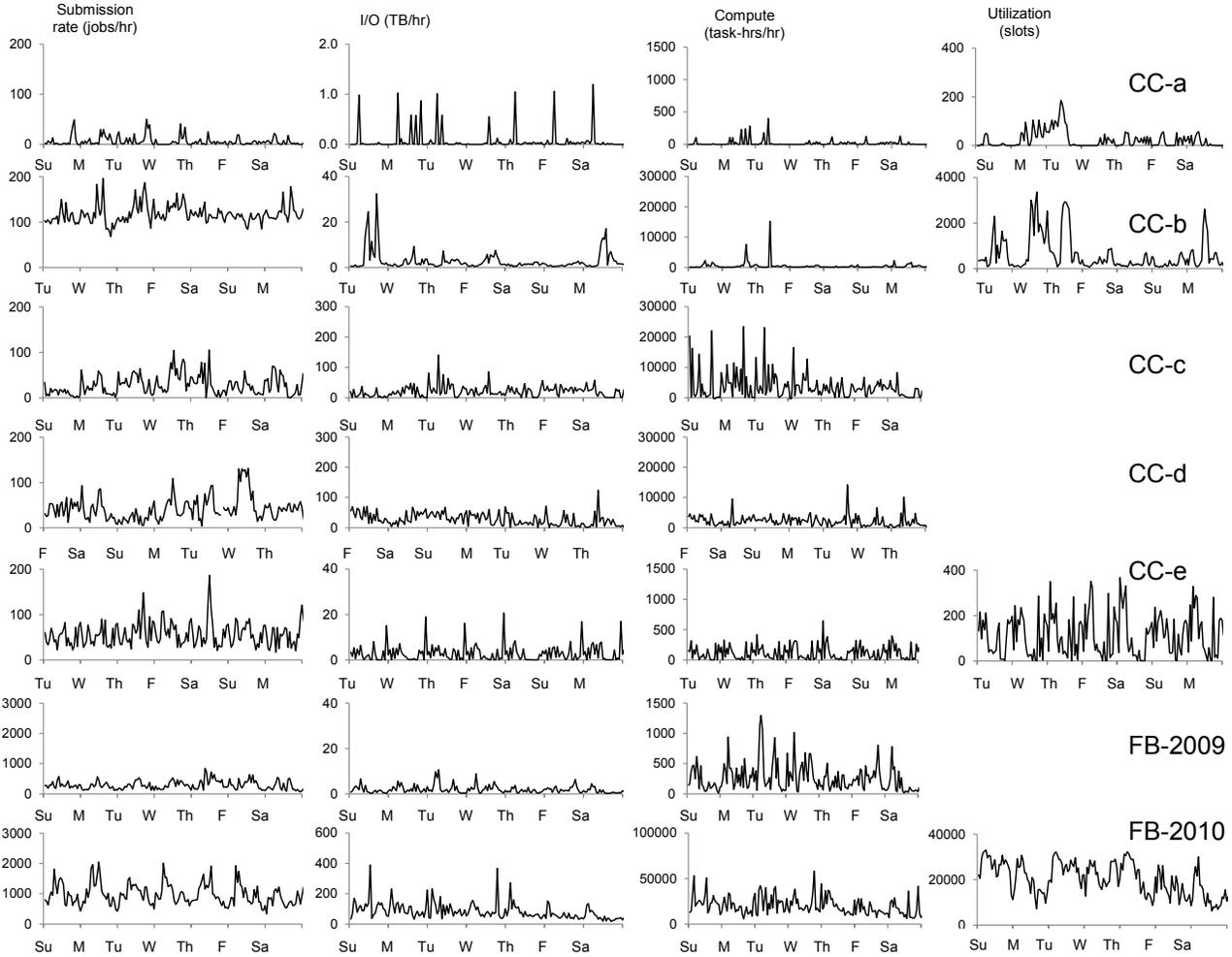

**Figure 7: Workload behavior over a week.** From left to right: (1) Jobs submitted per hour. (2) Aggregate I/O (i.e., input + shuffle + output) size of jobs submitted. (3) Aggregate map and reduce task time in task-hours of jobs submitted. (4) Cluster utilization in average active slots. From top row to bottom, showing `CC-a`, `CC-b`, `CC-c`, `CC-d`, `CC-e`, `FB-2009`, and `FB-2010` workloads. Note that for `CC-c`, `CC-d`, and `FB-2009`, the utilization data is not available from the traces. Also note that some time axes are misaligned due to short, week-long trace lengths (`CC-b` and `CC-e`), or gaps from missing data in the trace (`CC-d`).

Figure 7 offers visual evidence to indicate the diversity of MapReduce workloads. There is significant variation in the shape of the graphs for both different dimensions of the same workloads (rows) and for the same workload dimension across different workloads (columns). Consequently, for cluster management problems that involve workload variation over time scales, such as load scheduling, load shifting, resource allocation, or capacity planning, approaches designed for one workload may be suboptimal or even counter-productive for another. As MapReduce use cases diversify and increase in scale, it becomes vital to develop workload management techniques that can target each specific workload.

### 5.2 Burstiness

Figure 7 also reveals bursty submission patterns across various dimensions. Burstiness is an often discussed prop-erty of time-varying signals, but it is often not precisely measured. One common way to attempt to measure it to use the peak-to-average ratio. There are also domain-specific metrics, such as for bursty packet loss on wireless links [47]. Here, we extend the concept of peak-to-average ratio to quantify burstiness.

We start defining burstiness first by using the median rather than the arithmetic mean as the measure of "average". Median is statistically robust against data outliers, i.e., extreme but rare bursts [30]. For two given workloads with the same median load, the one with higher peaks, that is, a higher peak-to-median ratio, is more bursty. We then observe that the peak-to-median ratio is the same as the $100^{th}$-percentile-to-median ratio. While the median is statistically robust to outliers, the $100^{th}$-percentile is not. This implies that the $99^{th}$, $95^{th}$, or $90^{th}$-percentile should also be calculated. We extend this line of thought and compute the general $n^{th}$-percentile-to-median ratio for a workload. We can graph this vector of values, with $\frac{n^{th}-percentile}{median}$ on the x-axis, versus $n$ on the y-axis. The resultant graph can be interpreted as a cumulative distribution of arrival rates per time unit, normalized by the median arrival rate. This graph is an indication of how bursty the time series is. A more horizontal line corresponds to a more bursty workload; a vertical line represents a workload with a constant arrival rate.

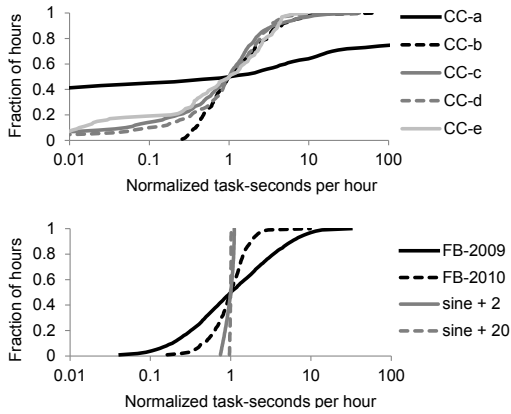

Figure 8: Workload burstiness. Showing cummulative distribution of task-time (sum of map time and reduce time) per hour. To allow comparison between workloads, all values have been normalized by the median task-time per hour for each workload. For comparison, we also show burstiness for artificial sine submit patterns, scaled with min-max range the same as mean (sine + 2) and 10% of mean (sine + 20).

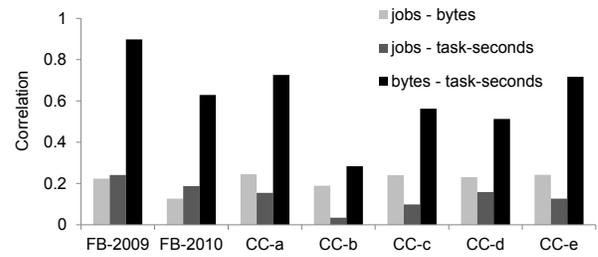

Figure 9: Correlation between different submission pattern time series. Showing pair-wise correlation between jobs per hour, (input + shuffle + output) bytes per hour, and (map + reduce) task times per hour.

Figure 8 graphs this metric for one of the dimensions of our workloads. We also graph two different sinusoidal signals to illustrate how common signals appear under this burstiness metric. Figure 8 shows that for all workloads, the highest and lowest submission rates are orders of magnitude from the median rate. This indicates a level of burstiness far above the workloads examined by prior work, which have more regular diurnal patterns [38, 48]. For the workloads here, scheduling and task placement policies will be essential under high load. Conversely, mechanisms for conserving energy will be beneficial during periods of low utilization.

For the Facebook workloads, over a year, the peak-to-median-ratio dropped from 31:1 to 9:1, accompanied by more internal organizations adopting MapReduce. This shows that multiplexing many workloads (workloads from many organizations) help decrease bustiness. However, the workload remains bursty.

## 5.3 Time Series Correlations

We also computed the correlation between the workload submission time series in all three dimensions. Specifically, we compute three correlation values: between the time-varying vectors $jobsSubmitted(t)$ and $dataSizeBytes(t)$, between $jobsSubmitted(t)$ and $computeTimeTaskSeconds(t)$, and between $dataSizeBytes(t)$ and $computeTimeTaskSeconds(t)$, where $t$ represents time in hourly granularity, and ranges over the entire trace duration.

The results are in Figure 9. The average temporal correlation between job submit and data size is 0.21; for job submit and compute time it is 0.14; for data size and compute time it is 0.62. The correlation between data size and compute time is by far the strongest. We can visually verify this by the $2^{nd}$ and $3^{rd}$ columns for CC-e in Figure 9. This indicates that MapReduce workloads remain data-centric rather than compute-centric. Also, schedulers and load balancers need to consider dimensions beyond number of active jobs.

Combined, the observations in this section mean that maximum jobs per second is the wrong performance metric to evaluate these systems. The nature of any workload bursts depends on the complex aggregate of data and compute needs of active jobs at the time, as well as the scheduling, placement, and other workload management decisions that determine how quickly jobs drain from the system. Any efforts to develop a TPC-like benchmark for MapReduce should consider a range of performance metrics, and stressing the system under realistic, multi-dimensional variations in workload intensity.

## 6. COMPUTATION PATTERNS

Previous sections looked at data and temporal patterns in the workload. As computation is an equally important aspect of MapReduce, this section identifies what are the common computation patterns for each workload. Specifically, we answer questions related to optimizing query-like programming frameworks:

— What % of cluster load come from these frameworks?
— What are the common uses of each framework?

We also answer questions with regard to job-level scheduling and execution planning:

— What are the common job types?
— What are the size, shape, and duration of these jobs?
— How frequently does each job type appear?

In traditional RDBMS, one can quantify query types by the operator (e.g. join, select), and the cardinality of the data processed for a particular query. Each operator can be characterized to consume a certain amount of resources based on the cardinality of the data they process. The analog to operators for MapReduce jobs are the map and reduce steps, and the cardinality of the data is quantified in our analysis by the number of bytes of data for the map input, intermediate shuffle, and reduce output stages.

We consider two complementary ways of grouping MapReduce jobs: (1) By the job name strings submitted to MapReduce, which gives us insights on the use of native MapReduce versus query-like programatic frameworks on top of MapReduce. For some frameworks, this analysis also reveals the frequency of the particular query-like operators that are used (§ 6.1). (2) By the multi-dimensional job description according to per-job data sizes, duration, and task times, which serve as a proxy for proprietary code, and indicate the size, shape, and duration of each job type (§ 6.2).

## 6.1  By Job Names

Job names are user-supplied strings recorded by MapReduce. Some computation frameworks built on top of MapReduce, such as Hive [1], Pig [3], and Oozie [2] generate the job names automatically. MapReduce does not currently impose any structure on job names. To simplify analysis, we focus on the first word of job names, ignoring any capitalization, numbers, or other symbols.

Figure 10 shows the most frequent first words in job names for each workload, weighted by number of jobs, the amount of I/O, and task-time. The FB-2010 trace does not have this information. The top figure shows that the top handful of words account for a dominant majority of jobs. When these names are weighted by I/O, Hive queries such as **insert** and other data-centric jobs such as data extractors dominate; when weighted by task-time, the pattern is similar, unsurprising given the correlation between I/O and task-time.

Figure 10 also implies that each workload consists of only a small number of common computation types. The reason is that job names are either automatically generated, or assigned by human operators using informal but common conventions. Thus, jobs with names that begin with the same word likely perform similar computation. The small number of computation types represent targets for static or even manual optimization. This will greatly simplify workload management problems, such as predicting job duration or resource use, and optimizing scheduling, placement, or task granularity.

Each workload services only a small number of MapReduce frameworks: Hive, Pig, Oozie, or similar layers on top of MapReduce. Figure 10 shows that for all workloads, two frameworks account for a dominant majority of jobs. There is ongoing research to achieve well-behaved multiplexing between different frameworks [32]. The data here suggests that multiplexing between two or three frameworks already covers the majority of jobs in all workloads. We believe this observation will remain valid in the future. As new frameworks develop, enterprise MapReduce users are likely to converge on an evolving but small set of mature frameworks for business critical computations.

Figure 10 also shows that for Hive in particular, **select** and **insert** form a large fraction of activity for several workloads. Only the FB-2009 workload contains a large fraction of Hive queries beginning with **from**. Unfortunately, this information is not available for Pig. Also, we see evidence of some direct migration of established RDBMS use cases, such as **etl** (Extract, Transform, Load) and **edw** (Enterprise Data Warehouse).

This information gives us some idea with regard to good targets for query optimization. However, more direct information on query text at the Hive and Pig level will be even more beneficial. For workflow management frameworks such as Oozie, it will be benefitial to have UUIDs to identify jobs belonging to the same workflow. For native MapReduce jobs, it will be desirable for the job names to contain a uniform convention of pre- and postfixes such as dates, computation types, steps in multi-stage processing, etc.. Obtaining information at that level will help translate insights from multi-operator RDBMS query execution planning to optimize multi-job MapReduce workflows.

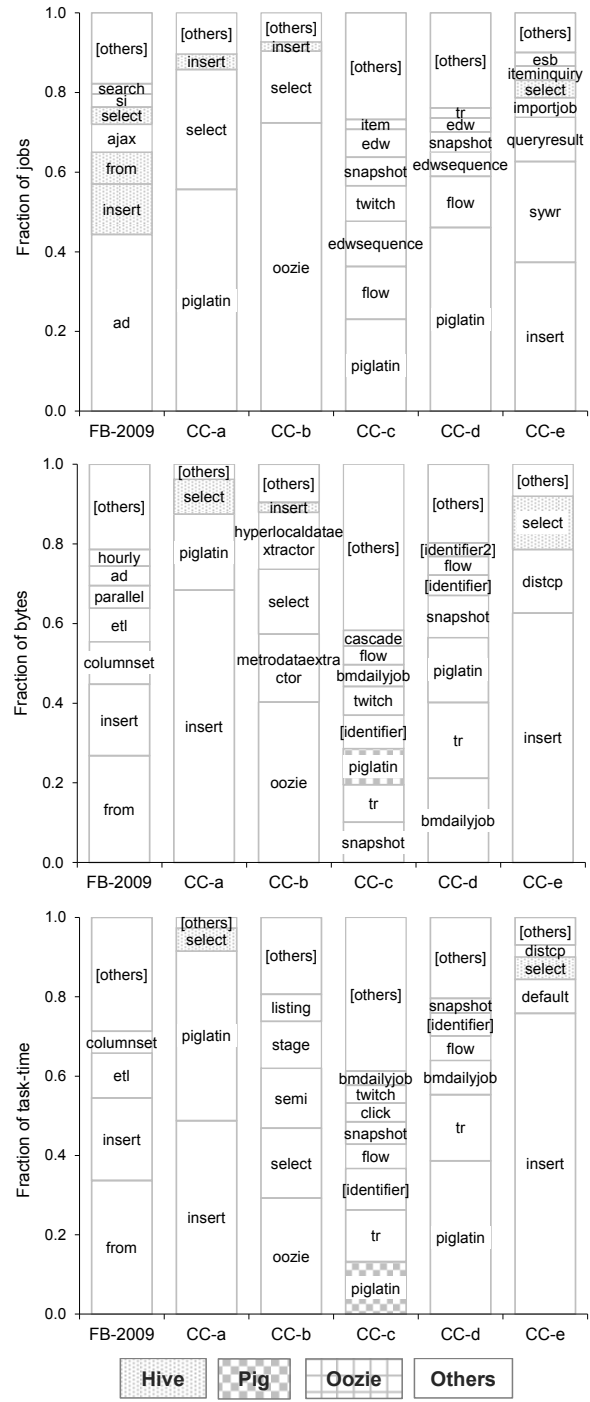

**Figure 10: The first word of job names for each workload, weighted by the number of jobs beginning with each word (top), total I/O in bytes (middle), and map/reduce task-time (bottom). For example, 44% of jobs in the FB-2009 workload have a name beginning with "ad", a further 12% begin with "insert"; 27% of all I/O and 34% of total task-time comes from jobs with names that begin with "from" (middle and bottom). The FB-2010 trace did not contain job names.**

## 6.2 By Multi-Dimensional Job Behavior

Another way to group jobs is by their multi-dimensional behavior. Each job can be represented as a six-dimensional vector described by input size, shuffle size, output size, job duration, map task time, and reduce task time. One way to group similarly behaving jobs is to find clusters of vectors close to each other in the six-dimensional space. We use a standard data clustering algorithm, k-means [9]. K-means enables quick analysis of a large number of data points and facilitates intuitive labeling and interpretation of cluster centers [17, 18, 41].

We use a standard technique to choose $k$, the number of job type clusters for each workload: increment $k$ until there is diminishing return in the decrease of intra-cluster variance, i.e., residual variance. Our previous work [17, 18] contains additional details of this methodology.

Table 2 summarizes our k-means analysis results. We have assigned labels using common terminology to describe the one or two data dimensions that separate job categories within a workload. A system optimizer would use the full numerical descriptions of cluster centroids.

We see that jobs touching <10GB of total data make up >92% of all jobs. These jobs are capable of achieving interactive latency for analysts, i.e., durations of less than a minute. The dominance of these jobs counters prior assumptions that MapReduce workloads consist of only jobs at TB scale and beyond. The observations validate research efforts to improve the scheduling time and the interactive capability of large-scale computing frameworks [14, 33, 39].

The dichotomy between very small and very large job has been identified previously for workload management of business intelligence queries [35]. Drawing on the lessons learned there, poor management of a single large job potentially impacts performance for a large number of small jobs.

The small-big job dichotomy implies that the cluster should be split into two tiers. There should be (1) a performance tier, which handles the interactive and semi-streaming computations and likely benefits from optimizations for interactive RDBMS systems, and (2) a capacity tier, which necessarily trades performance for efficiency in using storage and computational capacity. The capacity tier likely assumes batch-like semantics. One can view such a setup as analogous to multiplexing OLTP (interactive transactional) and OLAP (potentially batch analytical) workloads. It is important to operate both parts of the cluster while simultaneously achieving performance and efficiency goals.

The dominance of small jobs complicates efforts to rein in stragglers [10], tasks that execute significantly slower than other tasks in a job and delay job completion. Comparing the job duration and task time columns indicate that small jobs contain only a handful of small tasks, sometimes a single map task and a single reduce task. Having few comparable tasks makes it difficult to detect stragglers, and also blurs the definition of a straggler. If the only task of a job runs slowly, it becomes impossible to tell whether the task is inherently slow, or abnormally slow. The importance of stragglers as a problem also requires re-assessment. Any stragglers will seriously hamper jobs that have a single wave of tasks. However, if it is the case that stragglers occur randomly with a fixed probability, fewer tasks per job means only a few jobs would be affected. We do not yet know whether stragglers occur randomly.

Interestingly, map functions in some jobs aggregate data, reduce functions in other jobs expand data, and many jobs contain data transformations in either stage. Such data ratios reverse the original intuition behind map functions as expansions, i.e., "maps", and reduction functions as aggregates, i.e., "reduces" [22].

Also, map-only jobs appear in all but two workloads. They form 7% to 77% of all bytes, and 4% to 42% of all task times in their respective workloads. Some are Oozie launcher jobs and others are maintenance jobs that operate on very little data. Compared with other jobs, map-only jobs benefit less from datacenter networks optimized for shuffle patterns [7, 8, 20, 29].

Further, FB-2010 and CC-c both contain jobs that handle roughly the same amount of data as others, but take considerably longer to complete versus jobs in the same workload with comparable data sizes. FB-2010 contains a job type that consumes only 10s of GB of data, but requires days to complete (Map only transform, 3 days). These jobs have inherently low levels of parallelism, and cannot take advantage of parallelism on the cluster, even if spare capacity is available.

Comparing the FB-2009 and FB-2010 workloads in Table 2 shows that job types at Facebook changed significantly over one year. The small jobs remain, and several kinds of map-only jobs remain. However, the job profiles changed in several dimensions. Thus, for Facebook, any policy parameters need to be periodically revisited.

Combined, the analysis once again reveals the diversity across workloads. Even though small jobs dominate all seven workloads, they are "small" in different ways for each workload. Further, the breadth of job shape, size, and durations across workloads indicates that microbenchmarks of a handful of jobs capture only a small sliver of job activity, and a truly representative benchmark will need to involve a much larger range of job types.

## 7. TOWARDS A BIG DATA BENCHMARK

In light of the broad spectrum of industrial data presented in this paper, it is natural to ask what implications we can draw with regard to building a TPC-style benchmark for MapReduce and similar big data systems. The workloads here are sufficient to characterize an emerging class of MapReduce workloads for interactive and semi-streaming analysis. However, the diversity of behavior across the workloads we analyzed means we should be careful when deciding which aspects of this behavior are representative enough to include in a benchmark. Below, we discuss some challenges associated with building a TPC-style benchmark for MapReduce and other big data systems.

**Data generation.**

The range of data set sizes, skew in access frequency, and temporal locality in data access all affect system perfor-mance. A good benchmark should stress the system with realistic conditions in all these areas. Consequently, a benchmark needs to pre-generate data that accurately reflects the complex data access patterns of real life workloads.

| | # Jobs | Input | Shuffle | Output | Duration | Map time | Reduce time | Label |
|---|---|---|---|---|---|---|---|---|
| CC-a | 5525 | 51 MB | 0 | 3.9 MB | 39 sec | 33 | 0 | Small jobs |
| | 194 | 14 GB | 12 GB | 10 GB | 35 min | 65,100 | 15,410 | Transform |
| | 31 | 1.2 TB | 0 | 27 GB | 2 hrs 30 min | 437,615 | 0 | Map only, huge |
| | 9 | 273 GB | 185 GB | 21 MB | 4 hrs 30 min | 191,351 | 831,181 | Transform and aggregate |
| CC-b | 21210 | 4.6 GB | 0 | 4.7 KB | 23 sec | 11 | 0 | Small jobs |
| | 1565 | 41 GB | 10 GB | 2.1 GB | 4 min | 15,837 | 12,392 | Transform, small |
| | 165 | 123 GB | 43 GB | 13 GB | 6 min | 36,265 | 31,389 | Transform, medium |
| | 31 | 4.7 TB | 374 MB | 24 MB | 9 min | 876,786 | 705 | Aggregate and transform |
| | 3 | 600 GB | 1.6 GB | 550 MB | 6 hrs 45 min | 3,092,977 | 230,976 | Aggregate |
| CC-c | 19975 | 5.7 GB | 3.0 GB | 900 MB | 4 min | 10,933 | 6,586 | Small jobs |
| | 477 | 1.0 TB | 4.2 TB | 920 GB | 47 min | 1,927,432 | 462,070 | Transform, light reduce |
| | 246 | 887 GB | 57 GB | 22 MB | 4 hrs 14 min | 569,391 | 158,930 | Aggregate |
| | 197 | 1.1 TB | 3.7 TB | 3.7 TB | 53 min | 1,895,403 | 886,347 | Transform, heavy reduce |
| | 105 | 32 GB | 37 GB | 2.4 GB | 2 hrs 11 min | 14,865,972 | 36,9846 | Aggregate, large |
| | 23 | 3.7 TB | 562 GB | 37 GB | 17 hrs | 9,779,062 | 14,989,871 | Long jobs |
| | 7 | 220 TB | 18 GB | 2.8 GB | 5 hrs 15 min | 66,839,710 | 758,957 | Aggregate, huge |
| CC-d | 12736 | 3.1 GB | 753 MB | 231 MB | 67 sec | 7,376 | 5,085 | Small jobs |
| | 214 | 633 GB | 2.9 TB | 332 GB | 11 min | 544,433 | 352,692 | Expand and aggregate |
| | 162 | 5.3 GB | 6.1 TB | 33 GB | 23 min | 2,011,911 | 910,673 | Transform and aggregate |
| | 128 | 1.0 TB | 6.2 TB | 6.7 TB | 20 min | 847,286 | 900,395 | Expand and Transform |
| | 43 | 17 GB | 4.0 GB | 1.7 GB | 36 min | 6,259,747 | 7,067 | Aggregate |
| CC-e | 10243 | 8.1 MB | 0 | 970 KB | 18 sec | 15 | 0 | Small jobs |
| | 452 | 166 GB | 180 GB | 118 GB | 31 min | 35,606 | 38,194 | Transform, large |
| | 68 | 543 GB | 502 GB | 166 GB | 2 hrs | 115,077 | 108,745 | Transform, very large |
| | 20 | 3.0 TB | 0 | 200 B | 5 min | 137,077 | 0 | Map only summary |
| | 7 | 6.7 TB | 2.3 GB | 6.7 TB | 3 hrs 47 min | 335,807 | 0 | Map only transform |
| FB-2009 | 1081918 | 21 KB | 0 | 871 KB | 32 s | 20 | 0 | Small jobs |
| | 37038 | 381 KB | 0 | 1.9 GB | 21 min | 6,079 | 0 | Load data, fast |
| | 2070 | 10 KB | 0 | 4.2 GB | 1 hr 50 min | 26,321 | 0 | Load data, slow |
| | 602 | 405 KB | 0 | 447 GB | 1 hr 10 min | 66,657 | 0 | Load data, large |
| | 180 | 446 KB | 0 | 1.1 TB | 5 hrs 5 min | 125,662 | 0 | Load data, huge |
| | 6035 | 230 GB | 8.8 GB | 491 MB | 15 min | 104,338 | 66,760 | Aggregate, fast |
| | 379 | 1.9 TB | 502 MB | 2.6 GB | 30 min | 348,942 | 76,736 | Aggregate and expand |
| | 159 | 418 GB | 2.5 TB | 45 GB | 1 hr 25 min | 1,076,089 | 974,395 | Expand and aggregate |
| | 793 | 255 GB | 788 GB | 1.6 GB | 35 min | 384,562 | 338,050 | Data transform |
| | 19 | 7.6 TB | 51 GB | 104 KB | 55 min | 4,843,452 | 853,911 | Data summary |
| FB-2010 | 1145663 | 6.9 MB | 600 B | 60 KB | 1 min | 48 | 34 | Small jobs |
| | 7911 | 50 GB | 0 | 61 GB | 8 hrs | 60,664 | 0 | Map only transform, 8 hrs |
| | 779 | 3.6 TB | 0 | 4.4 TB | 45 min | 3,081,710 | 0 | Map only transform, 45 min |
| | 670 | 2.1 TB | 0 | 2.7 GB | 1 hr 20 min | 9,457,592 | 0 | Map only aggregate |
| | 104 | 35 GB | 0 | 3.5 GB | 3 days | 198,436 | 0 | Map only transform, 3 days |
| | 11491 | 1.5 TB | 30 GB | 2.2 GB | 30 min | 1,112,765 | 387,191 | Aggregate |
| | 1876 | 711 GB | 2.6 TB | 860 GB | 2 hrs | 1,618,792 | 2,056,439 | Transform, 2 hrs |
| | 454 | 9.0 TB | 1.5 TB | 1.2 TB | 1 hr | 1,795,682 | 818,344 | Aggregate and transform |
| | 169 | 2.7 TB | 12 TB | 260 GB | 2 hrs 7 min | 2,862,726 | 3,091,678 | Expand and aggregate |
| | 67 | 630 GB | 1.2 TB | 140 GB | 18 hrs | 1,545,220 | 18,144,174 | Transform, 18 hrs |

**Table 2: Job types in each workload as identified by k-means clustering, with cluster sizes, centers, and labels. Map and reduce time are in task-seconds, i.e., a job with 2 map tasks of 10 seconds each has map time of 20 task-seconds. Note that the small jobs dominate all workloads.**

**Processing generation.**

The analysis in this paper reveals challenges in accurately generating a processing stream that reflects real life workloads. Such a processing stream needs to capture the size, shape, and sequence of jobs, as well as the aggregate cluster load variation over time. It is non-trivial to tease out the dependencies between various features of the processing stream, and even harder to understand which ones we can omit for a large range of performance comparison scenarios.

**Mixing MapReduce and query-like frameworks.**

The heavy use of query-like frameworks on top of MapReduce indicates that future cluster management systems need to efficiently multiplex jobs both written in the native MapReduce API, and from query-like frameworks such as Hive, Pig, and HBase. Thus, a representative benchmark also needs to include both types of processing, and multiplex them in realistic mixes.

**Scaled-down workloads.**

The sheer data size involved in the workloads means that it is economically challenging to reproduce workload behavior at production scale. One can scale down workloads proportional to cluster size. However, there are many ways to describe both cluster and workload size. One could normalize workload size parameters such as data size, number of jobs, or the processing per data, against cluster size parameters such as number of nodes, CPU capacity, or available memory. It is not clear yet what would be the best way to scale down a workload.

**Empirical models.**

The workload behaviors we observed do not fit any well-known statistical distributions (the single exception being Zipf distribution in data access frequency). It is necessary for a benchmark to assume an empirical model of workloads, i.e., the workload traces *are* the model. This is a departure from some existing TPC-\* benchmarking approaches, where the targeted workload are such that some simple models can be used to generate data and the processing stream [51].

**A true workload perspective.**

The data in the paper indicate the shortcomings of microbenchmarks that execute a small number of jobs one at a time. They are useful for diagnosing subcomponents of a system subject to very specific processing needs. A big data benchmark should assume the perspective already reflected in TPC-\* [50], and treat a workload as a steady processing stream involving the superposition of many processing types.

**Workload suites.**

The workloads we analyzed exhibit a wide range of behavior. If this diversity is preserved across more workloads, we would be compelled to accpet that no single set of behaviors are representative. In that case, we would need to identify as small suite of workload classes that cover a large range of behavior. The benchmark would then consist not of a single workload, but a workload suite. Systems could trade optimized performance for one workload type against more average performance for another.

**A stopgap tool.**

We have developed and deployed Statistical Workload Injector for MapReduce (`https://github.com/SWIMProject-UCB/SWIM/wiki`). This is a set of New BSD Licensed workload replay tools that partially address the above challenges. The tools can pre-populate HDFS using uniform synthetic data, scaled to the number of nodes in the cluster, and replay the workload using synthetic MapReduce jobs. The workload replay methodology is further discussed in [18]. The SWIM repository already includes scaled-down versions of the FB-2009 and FB-2010 workloads. Cloudera has allowed us to contact the end customers directly and seek permission to make public their traces. We hope the replay tools can act as a stop-gap while we progress towards a more thorough benchmark, and the workload repository can contribute to a scientific approach to designing big data systems such as MapReduce.

# 8. SUMMARY AND CONCLUSIONS

To summarize the analysis results, we directly answer the questions raised in Section 2.1. The observed behavior spans a wide range across workloads, as we detail below.

1. For optimizing the underlying storage system:
   - Skew in data accesses frequencies range between an 80-1 and 80-8 rule.
   - Temporal locality exists, and 80% of data re-accesses occur on the range of minutes to hours.

2. For workload-level provisioning and load shaping:
   - The cluster load is bursty and unpredictable.
   - Peak-to-median ratio in cluster load range from 9:1 to 260:1.

3. For job-level scheduling and execution planning:
   - All workloads contain a range of job types, with the most common being small jobs.
   - These jobs are small in all dimensions compared with other jobs in the same workload. They involve 10s of KB to GB of data, exhibit a range of data patterns between the map and reduce stages, and have durations of 10s of seconds to a few minutes.

   - The small jobs form over 90% of all jobs for all workloads. The other job types appear with a wide range of frequencies.

4. For optimizing query-like programming frameworks:

   - The cluster load that comes from these frameworks is up to 80% and at least 20%.
   - The frameworks are generally used for interactive data exploration and semi-streaming analysis. For Hive, the most commonly used operators are `insert` and `select`; `from` is frequently used in only one workload. Additional tracing at the Hive/Pig/HBase level is required.

5. For performance comparison between systems:
   - A wide variation in behavior exists between workloads, as the above data indicates.
   - There is sufficient diversity between workloads that we should be cautious in claiming any behavior as "typical". Additional workload studies are required.

The analysis in this paper has several repercussions: (1). MapReduce has evolved to the point where performance claims should be qualified with the underlying workload assumptions, e.g., by replaying a suite of workloads. (2). System engineers should regularly re-assess design priorities subject to evolving use cases. Prerequisites to these efforts are workload replay tools and a public workload repository, so that engineers can share insights across different enterprise MapReduce deployments.

Future work should seek to improve analysis and monitoring tools. Enterprise MapReduce monitoring tools [21] should perform workload analysis automatically, present graphical results in a dashboard, and ship only the anonymized and aggregated metrics for workload comparisons offsite. Most importantly, tracing capabilities at the Hive, Pig, and HBase level should be improved. An analysis of query text at that level will reveal further insights, and expedite translating RDBMS knowledge to optimize MapReduce and solve real life problems involving large-scale data.

Improved tools will facilitate the analysis of more workloads, over longer time periods, and for additional statistics. This improves the quality and generality of the derived design insights, and contributes to the overall efforts to identify common behavior. The data in this paper indicate that we need to look at a broader range of use cases before we can build a truly representative big data benchmark.

We invite cluster operators and the broader data management community to share additional knowledge about their MapReduce workloads. To contribute, retain the job history logs generated by existing Hadoop tools, run the tools at `https://github.com/SWIMProjectUCB/SWIM/wiki/Analyze-e-historical-cluster-traces-and-synthesize-representative-workload`, and share the results.

# 9. ACKNOWLEDGMENTS

The authors are grateful for the feedback from our colleagues at UC Berkeley AMP Lab, Cloudera, Facebook, and other industrial partners. We especially appreciate the inputs from Archana Ganapathi, Anthony Joseph, David Zats, Matei Zaharia, Jolly Chen, Todd Lipcon, Aaron T. Myers, John Wilkes, and Srikanth Kandula. This research is supported in part by AMP Lab (`https://amplab.cs.`

berkeley.edu/sponsors/), and the DARPA- and SRC-funded MuSyC FCRP Multiscale Systems Center.